\documentclass[twocolumn,showpacs,preprintnumbers,amsmath,amssymb]{revtex4}

\usepackage{graphicx}
\usepackage{dcolumn}
\usepackage{bm}

\begin{document}


\title{Four-Photon Interference with Asymmetric Beam Splitters}

\author{B. H. Liu, F. W. Sun,  Y. X. Gong, Y. F. Huang, and G. C. Guo}
 \affiliation{Key Laboratory of Quantum Information,
 University of Science and Technology of China, \\CAS, Hefei, 230026, the People's Republic of
 China}
\author{Z. Y. Ou$^*$}
 \affiliation{Department of Physics, Indiana University-Purdue University
Indianapolis, \\402 N. Blackford Street, Indianapolis, IN 46202,
USA }


\begin{abstract}
Two experiments of four-photon interference are performed with two
pairs of photons from parametric down-conversion with the help of
asymmetric beam splitters. The first experiment is a
generalization of the Hong-Ou-Mandel interference effect to two
pairs of photons while the second one utilizes this effect to
demonstrate a four-photon de Broglie wavelength of $\lambda/4$ by
projection measurement.
\end{abstract}

\maketitle

Multi-photon quantum interference plays a pivotal role in quantum
information sciences. Although two-photon interference has been
widely studied \cite{man} and is applied to some quantum
information protocols \cite{zei}, quantum interference of more
than two photons has only recently been the focus of research
because of its role in the fundamental study of quantum
nonlocality \cite{GHZ,pan} and the improvement in the precision
phase measurement \cite{bol,ou,bot,wal,mit}.

The most well-known two-photon interference is the Hong-Ou-Mandel
effect \cite{hom}, where two photons enter a symmetric beam
splitter (50:50) from two sides respectively. Because of
two-photon destructive interference, the probability for the two
photons to exit at separate ports is zero. However, generalization
to higher photon number is not straightforward. For example, for
an input state of $|2_a,2_b\rangle$, i.e., two from each side
respectively (Fig.1a) to a symmetric beam splitter, there is a
nonzero probability for the $|2_A,2_B\rangle$ state at the output
\cite{orw1}, contrary to the two-photon counterpart.

Recently, however, Wang and Kobayashi \cite{wan} proposed a
generalization of the Hong-Ou-Mandel effect to three photons with
an asymmetric beam splitter ($T\ne R$). With a state of
$|2_a,1_b\rangle$ input at a beam splitter with $T=2R=2/3$, the
probability is zero for the state $|2_A,1_B\rangle$ in the output
state, due to three-photon interference \cite{wan}.  Sanaka {\it
et al.} \cite{san} demonstrated experimentally this three-photon
Hong-Ou-Mandel effect.

Wang and Kobayashi went further and proposed a three-photon
interferometer, which shows a three-photon de Broglie wavelength.
The de Broglie wavelength of a multi-photon state is the
equivalent matter wavelength for all the photons as one entity
\cite{jac}. Thus the de Broglie wavelength for an $N$-photon state
is simply $\lambda/N$ with $\lambda$ as the single-photon
wavelength. It will show up in an interference fringe that is $N$
times finer than the regular single-photon fringe. It has
applications in precision phase measurement and lithography
\cite{bol,ou,bot,wal,mit}. Liu {\it et al.} \cite{liu} implemented
the scheme by Wang and Kobayashi and observed an interference
pattern with the three-photon de Broglie wavelength.

\begin{figure}
\centerline{
\includegraphics[width = 2.5in]{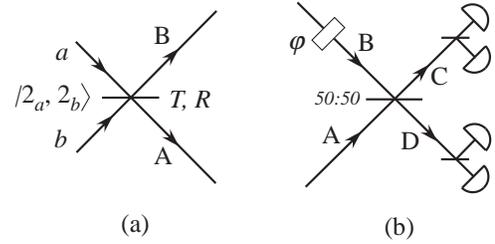}}
\caption{(a) Hong-Ou-Mandel interferometer with asymmetric beam
splitter and (b) formation of an interferometer for the de Broglie
wavelength of four photons.}
\end{figure}

In this Letter, we will generalize the idea of Wang and Kobayashi
to the four-photon case. We send the input state of
$|2_a,2_b\rangle$ to an asymmetric beam splitter and observe a
four-photon Hong-Ou-Mandel effect with proper adjustment of the
transmissivity of the beam splitter. Then we form an
interferometer and demonstrate the de Broglie wavelength for four
photons.

When a state of $|2_a,2_b\rangle$ enters an asymmetric beam
splitter with $T\ne R$ (Fig.1a), the output state is \cite{cam}
\begin{eqnarray}
&&|\Psi_4\rangle_{out} = \sqrt {6}TR \big(|4_A, 0_B\rangle + |0_A,
4_B\rangle\big)\cr &&\hskip .7in + \sqrt{6TR}(T-R)\big(|3_A,
1_B\rangle - |1_A, 3_B\rangle\big)\cr &&\hskip .8in +
\big[(T-R)^2-2TR\big]|2_A, 2_B\rangle.\label{3}
\end{eqnarray}
For a symmetric beam splitter with $T=R=1/2$, we find a nonzero
probability for the state $|2_A,2_B\rangle$ in the output, as we
discussed earlier. But when $(T-R)^2-2TR =0$ or
$T=(3\pm\sqrt{3})/6, R=(3\mp\sqrt{3})/6$, the $|2_A,2_B\rangle$
term disappears from Eq.(\ref{3}) and the probability of detecting
two photons at each side is zero, i.e., $P_4(2_A,2_B)=0$. Hence,
we realize a generalized Hong-Ou-Mandel effect for two pairs of
photons.

If we follow the outputs by another symmetric beam splitter as
shown in Fig.1b, the $|3_A,1_B\rangle$ and  $|1_A,3_B\rangle$
states in Eq.(\ref{3}) will not contribute to the probability
$P_4(2_C,2_D)$ of detecting four photons with two at each side due
to a two-photon Hong-Ou-Mandel effect. Since the $|2_A,2_B\rangle$
state in Eq.(\ref{3}) is cancelled out when $T=(3\pm\sqrt{3})/6,
R=(3\mp\sqrt{3})/6$, only the part of $|4, 0\rangle + |0,
4\rangle$ in Eq.(\ref{3}) will contribute to $P_4(2_C,2_D)$,
leading to a four-photon interference effect with
\begin{eqnarray}
P_4(2_C,2_D)\propto 1+\cos 4\varphi,\label{4}
\end{eqnarray}
where $\varphi$ is the single-photon phase difference between A
and B. This can be easily confirmed by evaluating the four-photon
detection probability $P_4(2_C,2_D) \propto \langle 2_a,2_b|\hat
C^{\dag 2}\hat D^{\dag 2} \hat D^2\hat C^2|2_a,2_b\rangle$ with
\begin{eqnarray}
\begin{cases}\hat C  = (\hat A + e^{j\varphi}\hat B)/\sqrt{2},\cr \hat D  =
(e^{j\varphi} \hat B - \hat A)/\sqrt{2},\end{cases} ~
\begin{cases}\hat A  = \sqrt{T}\hat a + \sqrt{R}\hat
b,\cr \hat B = \sqrt{T}\hat b - \sqrt{R}\hat a,\end{cases}
\label{5}
\end{eqnarray}
where $T=(3\pm\sqrt{3})/6, R=1-T$.
\begin{figure}
\centerline{
\includegraphics[width = 3in]{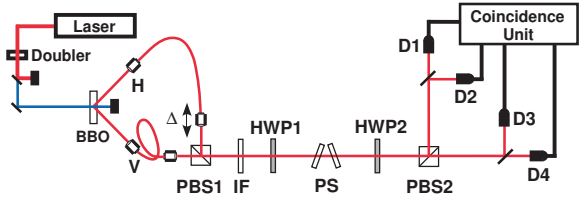}}
\caption{The layout of the experiment. PBS: polarization beam
splitter; HWP: half wave plate; PS: phase shifter; IF:
interference filter; D1-D4: photo-detectors.}
\end{figure}

Experimental implementation is shown in Fig.2, where the
four-photon state of $|2_a,2_b\rangle$ is produced from a type-II
parametric down-conversion process pumped by 150 fsec
frequency-doubled pulses from a mode-locked Ti:Sapphire laser
operating at 780 nm. The 2mm long $\beta$-Barium Borate (BBO)
crystal is so oriented that it produces two beam-like orthogonally
polarized fields at the degenerate wavelength of 780 nm
\cite{tak}. The horizontal (H) and the vertical (V) polarized
fields are first coupled into single-mode fibers and are
recombined with a polarization beam splitter (PBS1) into one beam
before passing through an interference filter (IF) of 3nm
bandpass. The filtered field is then feeded into the four-photon
interferometer of Fig.1 But the beam splitters of Fig.1 are
equivalently replaced by two polarization rotators (HWPs) and
another polarization beam splitter (PBS2). Thus it is a
polarization interferometer. A phase shifter (PS), made of two
synchronically rotating birefringent quartz plates, is inserted
between the two HWPs to introduce a variable single-photon phase
shift $\varphi$ between the two orthogonal polarizations. The
input four-photon state of $|2_H,2_V\rangle$ is generated via two
pairs of down-converted photons.

\begin{figure}
\centerline{
\includegraphics[width = 3in]{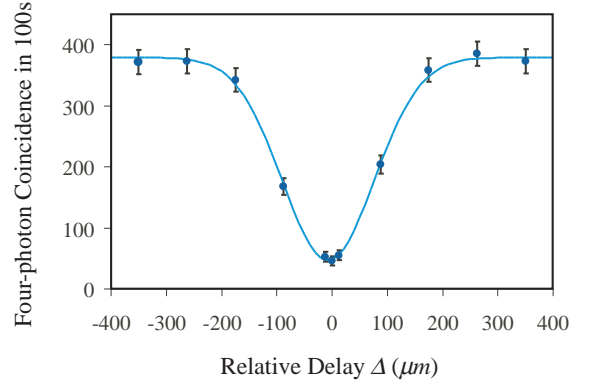}}
\caption{Four-photon coincidence as a function of the relative
delay $\Delta$ between the H- and V-polarizations.}
\end{figure}

\begin{figure}
\centerline{
\includegraphics[width = 3in]{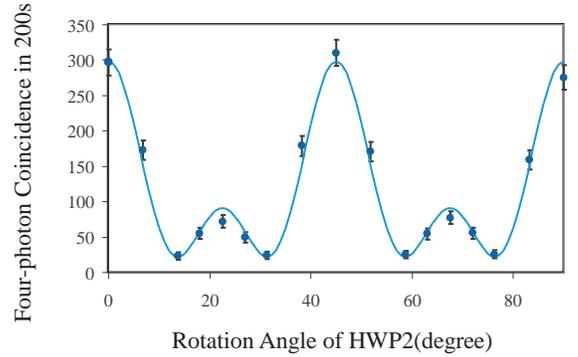}}
\caption{Four-photon coincidence as a function of the rotation
angle of HWP2.}
\end{figure}

In the first experiment, the rotation angle from HWP1 is set to
zero so that it has no effect on the H- and V-polarizations except
for a relative delay but that from HWP2 is set at $\theta =
13.7^{\circ}$ so that $\cos^22\theta =(3+\sqrt{3})/6$ (angle of
polarization rotation is $2\theta$). HWP2 and PBS2 together are
equivalent to an asymmetric beam splitter of $T=\cos^22\theta$ and
$R=\sin^22\theta$. The fiber coupler for the H-polarized photons
is mounted on a micro-translation stage to introduce a delay
$\Delta$ between the H- and V-polarizations. Four-photon
coincidence counts are registered to measure the probability
$P_4(2_C, 2_D)$ as a function of the delay between the H- and
V-polarizations. The data is shown in Fig.3 after background
subtraction. It shows the typical Hong-Ou-Mandel dip with a
visibility of 88\% and a full width at half height of 196$\mu m$,
which are derived from a least square fit to a mathematically
convenient Gaussian shape (the solid curve). The less than 100\%
visibility is a result of imperfect temporal mode match between
the two pairs of down-converted photons \cite{orw,tsu,ou1}.  To
verify that we indeed have the correct $T$ and $R$ with $\theta$
at $13.7^{\circ}$, we fix the delay $\Delta$ at the bottom of the
dip in Fig.3 but change $\theta$. The measured four-photon
coincidence counts after background subtraction are plotted as a
function of $\theta$ in Fig.4, which shows four minima at $\theta
= 13.7^{\circ}, 31.3^{\circ}, 58.7^{\circ}, 76.3^{\circ}$,
corresponding to $\cos^22\theta = T=(3\pm\sqrt{3})/6$. Again, the
minimum values do not go to zero, due to imperfect temporal mode
match. The solid curve is a least square fit to the function
\cite{ou2}
\begin{eqnarray}
&&P_4(\theta) = C\big[(1-1.5\sin^2 4\theta)^2 + (3\sin^2
4\theta-1)\cr&&\hskip 0.8in \times(1-\sin^2 4\theta)(1-{\cal
E/A})/2\big],\label{6}
\end{eqnarray}
where $C$ is a scaling factor and ${\cal E/A}$ is a parameter that
characterizes the temporal mismatch \cite{orw,ou1}. Note that when
${\cal E/A} = 1$, the function in Eq.(\ref{6}) touches zero at the
four minima, corresponding to the perfect mode match.

\begin{figure}
\centerline{
\includegraphics[width = 3in]{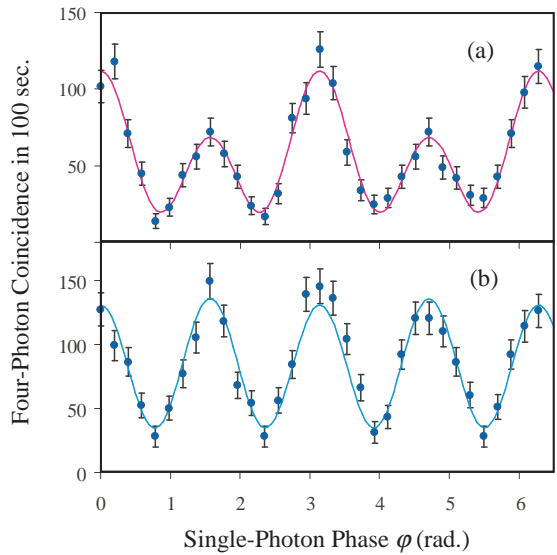}}
\caption{Four-photon coincidence as a function of the phase
difference $\varphi$ between H- and V-polarizations. HWP1 is set
at (a) $\theta=13.7^{\circ}$ and (b) $\theta=13.1^{\circ}$.}
\end{figure}

In the next experiment, we set the delay $\Delta$ at the bottom of
the dip in Fig.3 and turn HWP1 to $13.7^{\circ}$ and HWP2 to
$22.5^{\circ}$ so that HWP1 serves as the asymmetric beam splitter
(BS1) with $T=\cos^2 (2\times 13.7^{\circ}) = (3+\sqrt{3})/6$ and
HWP2 as the $50:50$ symmetric beam splitter (BS2) in Fig.1. We
measure the four-photon coincidence as a function of the
single-photon phase difference $\varphi$ between the H- and
V-polarizations via the phase shifter (PS in Fig.2). In this way,
we form a polarization interferometer equivalent to that in Fig.1.
The result of this measurement is shown in Fig.5a. Although it
reaches minimum at the values around $\varphi =\pi/4, 3\pi/4,
5\pi/4, 7\pi/4$, as predicted by Eq.(\ref{4}), the coincidence has
very uneven maxima. This is caused by the imperfect cancellation
of the $|2_A,2_B\rangle$ term in Eq.(\ref{3}) due to temporal mode
mismatch \cite{ou2} as shown in the nonzero minimum in Fig.3. The
existence of the $|2_A,2_B\rangle$ term will add a $\cos2\varphi$
term to Eq.(\ref{4}) resulting from interference between
$|4_A,0_B\rangle + |0_A,4_B\rangle$ and $|2_A,2_B\rangle$. The
data in Fig.5a fits very well to the function
\begin{eqnarray}
P_4(2_C,2_D)=  C( 1+ V_4\cos 4\varphi+V_2\cos 2\varphi) \label{7}
\end{eqnarray}
with $V_4=0.62$ and $V_2=0.39$.

Fortunately, the uneven peaks in Fig.5a can be balanced \cite{ou2}
by slightly adjusting HWP1 away from $13.7^{\circ}$ to
$13.1^{\circ}$, as shown in Fig.5b. The least square fit for the
data in Fig.5b to the function in Eq.(\ref{7}) gives $V_4=0.59$
and $V_2=-0.03$. The smallness of $V_2$ indicates a good
cancellation of the $\cos2\varphi$ term in Eq.(\ref{7}).

In conclusion, we demonstrated both the generalized Hong-Ou-Mandel
effect and the de Broglie wavelength of four photons with two
pairs of down-converted photons in a scheme involving asymmetric
beam splitters. These two effects are a result of four-photon
interference.

 This work was funded by National Fundamental
Research Program of China (2001CB309300), the Innovation funds
from Chinese Academy of Sciences, and National Natural Science
Foundation of China (Grant No. 60121503). ZYO is supported by the
US National Science Foundation under Grant No. 0245421 and No.
0427647.

$^*$Email: zou@iupui.edu

\begin{thebibliography}{99}
\bibitem{ou2} A more detailed multi-mode theory similar to
Refs.\cite{orw,ou1} will take into account the temporal mismatch
between the two pairs of down-converted photons. This theory will
be presented elsewhere.
\end{thebibliography}

\end{document}